\pdfoutput=1

\documentclass{article}
\usepackage{spconf,amsmath,graphicx}

\usepackage{tabu}                           
\usepackage{multirow,makecell}              

\usepackage{xspace}

\def\eg{\textit{e.g.}\xspace}
\def\ie{\textit{i.e.}\xspace}
\def\pa{presence\slash absence\xspace}
\def\Pa{Presence\slash absence\xspace}
\def\PA{Presence\slash Absence\xspace}

\usepackage[bibstyle=ieee,sorting=none,giveninits=true,citestyle=numeric]{biblatex}
\AtNextBibliography{\footnotesize}

\usepackage[pdftex,colorlinks]{hyperref}
\usepackage[all]{hypcap}    

\title{\vspace{-1.3em}CONNECTIONIST TEMPORAL LOCALIZATION FOR SOUND EVENT DETECTION WITH SEQUENTIAL LABELING}

\name{Yun Wang and Florian Metze
\thanks{\scriptsize This work was supported in part by a gift award from Robert Bosch LLC
and a faculty research award from Google.
It used the ``comet'' and ``bridges'' clusters of the
XSEDE environment \cite{XSEDE}, supported by NSF grant number ACI-1548562.}}
\address{Language Technologies Institute, Carnegie Mellon University, Pittsburgh, PA, U.S.A. \\
\texttt{maigoakisame@gmail.com, fmetze@cs.cmu.edu}}

\begin{document}
\ninept

\maketitle

\begin{abstract}
Research on sound event detection (SED) with weak labeling has mostly
focused on \pa labeling, which provides no temporal information
at all about the event occurrences.
In this paper, we consider SED with sequential labeling,
which specifies the temporal order of the event boundaries.
The conventional connectionist temporal classification (CTC) framework,
when applied to SED with sequential labeling,
does not localize long events well due to a ``peak clustering'' problem.
We adapt the CTC framework and propose
connectionist temporal localization (CTL),
which successfully solves the problem.
Evaluation on a subset of Audio Set shows that
CTL closes a third of the gap between \pa labeling
and strong labeling, demonstrating the usefulness of the extra
temporal information in sequential labeling.
CTL also makes it easy to combine sequential labeling
with \pa labeling and strong labeling.
\end{abstract}


\section{Introduction}
\label{sec:intro}

Sound event detection (SED) is the task of classifying and localizing
occurrences of sound events in audio streams.
The training of SED models used to rely upon \emph{strong labeling},
which specifies the type, onset time and offset time of each sound event occurrence.
Such labeling, however, is very tedious to obtain by hand.
In order to scale SED up, many successful attempts have been made
to train SED systems with \emph{weak labeling},
such as~\cite{hershey2017cnn, kong2018audio, yu2018multi}
and our TALNet~\cite{TALNet-arXiv}.
Even though trained with weak labeling, some of these systems
are able to temporally localize events in their output.

When the term \emph{weak labeling} is used in the literature,
it often specifically refers to \emph{\pa labeling},
which only specifies the types of sound events present in a recording
but does not provide any temporal information.
\Pa labeling is popular because it takes the least effort to produce;
as such, Audio Set~\cite{AudioSet}, the currently largest corpus for SED,
is also labeled this way.
In this paper, however, we study SED with sequential labeling,
which specifies the order of the boundaries of events occurring
in each recording.
We demonstrate that the extra temporal information in sequential labeling,
though incomplete, can still improve the localization of sound events.

Connectionist temporal classification (CTC)~\cite{CTC} is a popular framework
used for speech recognition when the supervision is sequential,
\eg phoneme sequences without temporal alignment~\cite{graves2014towards}.
CTC has been directly applied to SED with sequential labeling in~\cite{hou2018polyphonic}
and a previous work of ours~\cite{Yun-ICASSP2017};
the latter found that a ``peak clustering'' problem
impeded the accurate localization of long sound events.
In this paper, we make three major modifications to CTC and propose
a connectionist temporal localization (CTL) framework,
which successfully solves the peak clustering problem.
Evaluation on a subset of Audio Set shows that
CTL closes a third of the gap between presence/absence labeling
and strong labeling.

Our CTL framework also provides a way to easily combine
multiple types of labeling, such as \pa labeling, sequential labeling,
and strong labeling.
When we have stronger labeling available in a smaller amount
and weaker labeling available in a larger amount,
such a combination makes it possible to fully exploit
the information in all the data.

\section{CTL: Motivation and Algorithm}
\label{sec:theory}

\subsection{Sequential Labeling}
\label{sec:seq-label}

In speech recognition, a typical form of supervision
is a phoneme sequence for each utterance without temporal alignment.
A direct analogy for SED would be a sequence of sound events
for each recording, but the order of sound events
can be hard to define when they overlap.
To avoid this problem, we define sequential labeling to be
a sequence of event boundaries.
For example, if the content of a recording can be described as
``a dog barks while a car passes by'',
the sequence of event boundaries will be:
car onset, dog onset, dog offset, car offset.
We denote this by \texttt{\underline{\'C\'D\`D\`C}}:
letters with the rising accent \texttt{\'C}, \texttt{\'D}
stand for the onsets of the ``car'' and ``dog'' events,
while letters with the falling accent \texttt{\`C} and \texttt{\`D}
stands for their offsets;
the underline means this is a sequence without temporal alignment.

For annotators, sequential labeling is not too much harder
to produce than \pa labeling; the difficulty mainly arises
when sound events occur densely or overlap.
In any case, it is still easier to produce than strong labeling,
because it is not necessary to mark the precise onset and offset times
of each sound event occurrence.
Also, sequential labeling may be automatically mined from
textual descriptions of audio recordings, such as
``a dog barks while a car passes by''.

\subsection{The Peak Clustering Problem of CTC}
\label{sec:peak-cluster}

CTC can be applied to SED with sequential labeling as follows.
First, we define the vocabulary of CTC output to include
the onset and offset labels of each event type,
plus a ``blank'' label (denoted by \texttt{-}).
For an SED system that deals with $n$ types of events,
the vocabulary size is $2n+1$.
A neural network (often with a recurrent layer)
predicts the frame-wise probability of each label in the vocabulary;
these probabilities sum to 1 at each frame.
The probabilities of specific temporal alignments
(\eg \texttt{-\'C\'D\`D\`C-}, \texttt{\'C-\'D\`D\`C\`C}) can be calculated
by multiplying the probabilities of individual labels at each frame.
The total probability of the ground truth sequence
(\eg \texttt{\underline{\'C\'D\`D\`C}})
is defined as the sum of the probabilities of all alignments
that can be reduced to the ground truth sequence by a
many-to-one mapping $\mathcal{B}$;
this mapping first collapses all consecutive repeating labels
into a single one, then removes all blank labels.
For example, both the alignments \texttt{-\'C\'D\`D\`C-} and \texttt{\'C-\'D\`D\`C\`C}
can be reduced to the unaligned sequence \texttt{\underline{\'C\'D\`D\`C}},
therefore $P(\texttt{\underline{\'C\'D\`D\`C}}) =
P(\texttt{-\'C\'D\`D\`C-}) + P(\texttt{\'C-\'D\`D\`C\`C})$
plus the probabilities of many other alignments.
A systematic forward algorithm is proposed in~\cite{CTC}
to compute this total probability efficiently.
The loss function for this recording is defined as
$-\log P(\texttt{\underline{\'C\'D\`D\`C}})$;
this can be minimized with any neural network training algorithm,
such as gradient descent.

When CTC is directly applied to SED with sequential labeling,
it has been found in~\cite{Yun-ICASSP2017} to detect short events well:
a peak appears in the frame-wise probabilities of the onset and offset
labels around the actual occurrence of the event.
For long events, however, CTC tends to predict peaks
for the onset and the offset next to each other,
which means the event is not well localized
(see Sec.~\ref{sec:ctl-result} for an example).

This ``peak clustering'' problem occurs for several reasons.
First, because sound events do not overlap too often,
adjacent onset and offset labels are an extremely common pattern
in the training label sequences.
As a result, CTC may misunderstand a pair of onset and offset labels as
collectively indicating the existence of an event,
instead of understanding them as separately indicating the event boundaries.
Second, the CTC loss function only mandates the order of the predicted labels,
without imposing any temporal constraints.
In this case, the recurrent layer of the network will prefer
to emit onset and offset labels next to each other,
because this minimizes the effort of memory.
The root cause of the ``peak clustering'' problem is that the output layer
of the network is only trained to detect event \emph{boundaries};
it is expected to keep ``silent'' both when an event is inactive
and when an event is continuing, despite the potentially
huge differences in the acoustic features.
When the network predicts the onset and offset labels
of a long event occurrence next to each other,
it actually does not violate this expectation on too many frames,
and does not have enough incentive to correct this behavior.

\subsection{Connectionist Temporal Localization}
\label{sec:ctl-algorithm}

In this section we make three major modifications to the CTC framework,
and present a connectionist temporal localization (CTL) framework
suitable for localizing sound events.
We also describe the corresponding forward algorithm for
calculating the total probability of an event boundary sequence.

The first modification addresses the root cause of the
``peak clustering'' problem: \emph{the output layer of the network
should predict the frame-wise probabilities of the events themselves
instead of those of the event boundaries}.
In this way, the network can learn to make different predictions
with different acoustic features.
The boundary probabilities are then derived from the event probabilities
using a ``rectified delta'' operator.
More formally, let $y_t(\texttt{E})$ be the probability of the event \texttt{E}
being active at frame $t$. Here $1 \le t \le T$,
where $T$ is the number of frames in the recording in question.
Let $z_t(\texttt{\'E})$ and $z_t(\texttt{\`E})$ be the probabilities
of the onset and offset labels of the event \texttt{E} at frame $t$.
We calculate them using the following equations:
\begin{equation}
\begin{aligned}
z_t(\texttt{\'E}) &= \max[0, y_t(\texttt{E}) - y_{t-1}(\texttt{E})] \\
z_t(\texttt{\`E}) &= \max[0, y_{t-1}(\texttt{E}) - y_t(\texttt{E})]
\end{aligned}
\label{eq:rectified-delta}
\end{equation}
In these equations we allow $t$ to range from 1 to $T+1$,
in order to accommodate events that start at the first frame or end at the last frame.
When $y_0(\texttt{E})$ or $y_{T+1}(\texttt{E})$ is referenced, we assume it to be 0.

Now we have the frame-wise probabilities of all event boundaries,
we only need to define the frame-wise probability of the blank.
However, a difficulty arises because the sum of the boundary probabilities
at a given frame may exceed 1.
To solve this problem, we make the second modification to CTC:
\emph{we treat the probabilities of different event boundaries at the same frame
as mutually independent, instead of mutually exclusive}.
In this way, the probability of no event boundaries occurring at frame $t$
can be calculated by:
\begin{equation}
\epsilon_t = \prod\nolimits_l [1 - z_t(l)]
\end{equation}
where $l$ goes over all event boundaries.
The probability of emitting a single event boundary $l$ at frame $t$ is then:
\begin{equation}
p_t(l) = z_t(l) \cdot \prod\nolimits_{l' \ne l} [1 - z_t(l')]
\label{eq:ptl}
\end{equation}
If we define
\begin{equation}
\delta_t(l) = \frac{z_t(l)}{1 - z_t(l)}
\end{equation}
Then Eq.~\ref{eq:ptl} reduces to
\begin{equation}
p_t(l) = \epsilon_t \cdot \delta_t(l)
\end{equation}

The assumption that boundary labels at the same frame are mutually independent
seems to eliminate the need for the blank label.
Indeed, the blank label in CTC serves two purposes:
(1) to allow emitting nothing at a frame, and
(2) to separate consecutive repetitions of the same label.
With the independence assumption, the first purpose is naturally achieved.
Here we make the third modification to CTC:
\emph{the mapping $\mathcal{B}$ no longer collapses
consecutive repeating labels into a single one}.
With this simplification, the blank label can be removed altogether.

The independence assumption also allows us to assess the probability
of emitting multiple labels at the same frame,
which is not possible with the standard CTC.
The probability of emitting multiple labels $l_1, \ldots, l_k$
together at frame $t$ can be calculated as
\begin{align}
p_t(l_1, \ldots, l_k) &= \prod\nolimits_{i=1}^k z_t(l_i) \cdot \prod\nolimits_{l \notin \{l_1, \ldots, l_k\}} [1 - z_t(l)] \nonumber \\
&= \epsilon_t \cdot \prod\nolimits_{i=1}^k \delta_t(l_i)
\label{eq:ptl_multi}
\end{align}

Now we can formulate our CTL forward algorithm.
What we want to find is the total probability of emitting
the ground truth label sequence $L = \underline{l_1, \ldots, l_{|L|}}$,
regardless of the temporal alignment.
What we are given is the frame-level probabilities of events $y_t(\texttt{E})$,
from which we can derive the probability $p_t(\cdot)$
of emitting zero, one or more labels at each frame by Eq.~\ref{eq:ptl_multi}.
Let $\alpha_t(i)$ be the probability of
having emitted exactly the first $i$ labels of $L$ after $t$ frames.
The $\alpha$'s can be computed with the following recurrence formula:
\begin{align}
\alpha_t(i) &= \sum\nolimits_{j=0}^i \alpha_{t-1}(i-j) \cdot p_t(l_{i-j+1}, \ldots, l_i) \nonumber \\
&= \sum\nolimits_{j=0}^i \alpha_{t-1}(i-j) \cdot \epsilon_t \cdot \prod\nolimits_{k=i-j+1}^i \delta_t(l_k)
\label{eq:ctl-trans}
\end{align}
In the summation, the index $j$ stands for the number of labels emitted at frame $t$.
The initial values are:
\begin{equation}
\alpha_0(i) = \left\{ \begin{array}{ll}
    1, & \quad \text{if}\ i = 0 \\
    0, & \quad \text{if}\ i > 0
\end{array} \right.
\end{equation}
The final value, $\alpha_{T+1}(|L|)$, is the total probability
of emitting the label sequence $L$, and its negative logarithm is the contribution
of the recording in question to the loss function.

Eq.~\ref{eq:ctl-trans} allows emitting arbitrarily many labels at the same frame.
When the ground truth label sequence is long, this can pose a problem of time complexity.
In practice, it is rare for multiple labels to be emitted at the same frame.
Therefore, it can be desirable to limit the maximum number of concurrent labels,
\ie the maximum value of $j$ in Eq.~\ref{eq:ctl-trans}.
We call this maximum value the \emph{max concurrence}.

\begin{figure*}[t]
\centering
\includegraphics[width=0.98\textwidth]{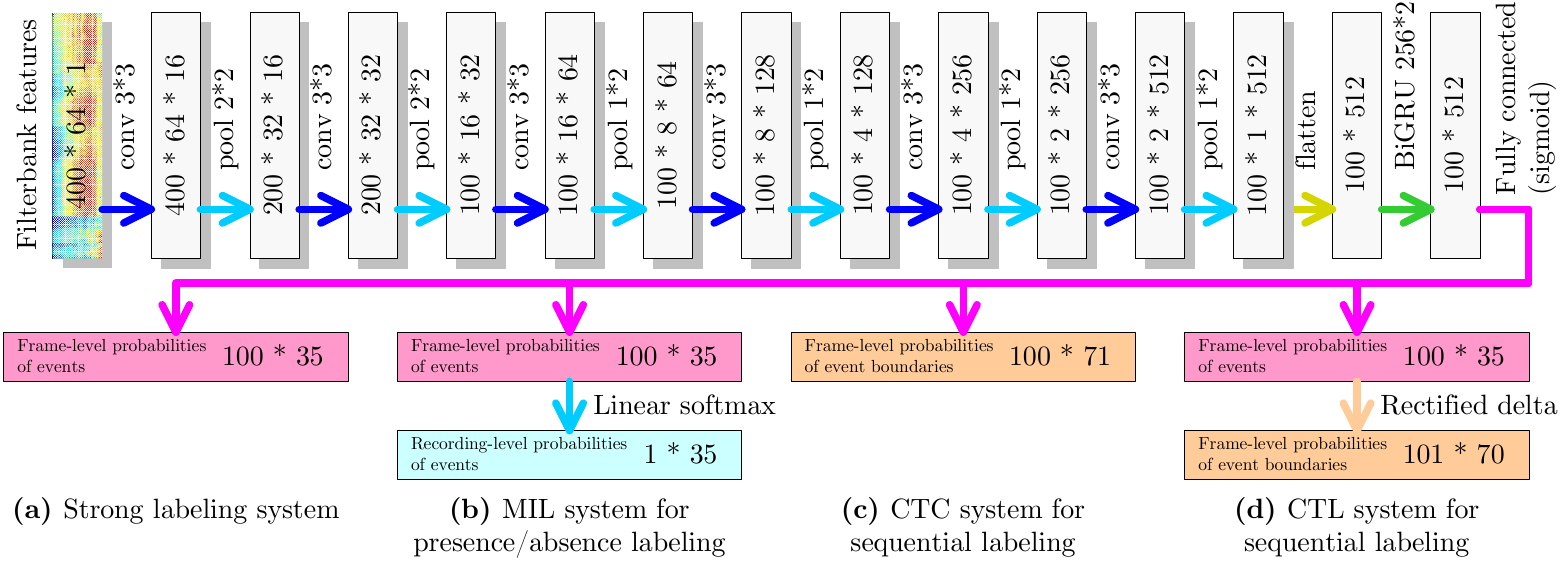}
\caption{Structures of the four networks trained in Sec.~\ref{sec:network}.
The shape is specified as ``frames * frequency bins * feature maps''
for \mbox{3-D} tensors (shaded), and
``frames * feature maps'' for \mbox{2-D} tensors.
``conv $n$*$m$'' stands for a convolutional layer with the specified
kernel size and ReLU activation;
batch normalization is applied before the ReLU activation.
``pool $n$*$m$'' stands for a max pooling layer with the specified stride.
\vspace{-0.5em}}
\label{fig:structure}
\end{figure*}

\section{Experiments}
\label{sec:exp}

\subsection{Data Preparation}
\label{sec:data}

We carried out experiments on a subset of Audio Set~\cite{AudioSet}.
Audio Set consists of over 2~million 10-second excerpts of
YouTube videos, labeled with the \pa of 527~types of sound events.
Because we would need sequential labeling for training and
strong labeling for evaluation, we generated sequential and strong
labeling for all the recordings using TALNet~\cite{TALNet-arXiv}
-- a state-of-the-art network trained with \pa labeling
that is good at localizing sound events.
We used a frame length of 0.1~s, so each recording consisted of 100~frames.

Not all of the 527~sound events types of Audio Set were labeled
with high quality, and the labels generated by TALNet would be even noisier.
To reduce the effect of such label noise,
we selected 35~sound event types that had relatively reliable labels
(see Table~4.1 of~\cite{Yun-PhD-Thesis} for a complete list).
Four of these event types (\texttt{speech}, \texttt{sing},
\texttt{music} and \texttt{crowd}) were overwhelmingly frequent;
we filtered the recordings of Audio Set to retain only those
that contained at least one of the remaining 31~types of sound events.
This left us with 359,741 training recordings, 4,879 validation recordings
and 5,301 evaluation recordings. The total duration of these recordings
is around 1,000 hours, or 18\% of entire corpus.

\subsection{Network Structures and Training}
\label{sec:network}

We trained four networks whose structures are illustrated in
Fig.~\ref{fig:structure}.
All the layers up to the GRU layer are shared across the four networks;
these layers highly resemble the hidden layers of TALNet~\cite{TALNet-arXiv},
but are shallower and narrower.
The four systems have different output ends.
The first system predicts the probabilities of the 35~types of sound events,
and directly receives strong labeling as supervision.
The second system is a multiple instance learning (MIL) system for \pa labeling:
it first predicts frame-wise probabilities, then aggregates them
into recording-level probabilities with a linear softmax pooling function
just like TALNet.
These two systems serve as the topline and the baseline
for the CTC and CTL systems.
The CTC system directly predicts the frame-wise probabilities
of event boundaries and the blank label;
the output layer has $35 * 2 + 1 = 71$ units.
The CTL system predicts the frame-wise probabilities of the events
and then derives the boundary probabilities with the ``rectified delta'' operator.
We tried max concurrence values of 1, 2 and 3.

The systems were trained using the Adam optimizer~\cite{Adam}
with a constant learning rate of $10^{-3}$.
The batch size was 500~recordings.
We applied data balancing to ensure that each minibatch contained roughly
equal numbers of recordings of each event type.
After every 200~minibatches (called a \emph{checkpoint}),
we evaluated the network's localization performance using
the frame-level $F_1$ macro-averaged across the 35~event types.
For the strong labeling, MIL and CTL systems,
we first tuned class-specific thresholds to optimize
the frame-level $F_1$ of each event type on the validation data,
then applied them directly to the evaluation data.
For the CTC system, we picked the most probable label at each frame,
and marked each event as active between innermost matching
pairs of onset and offset labels.
\vspace{-0.3em}

\subsection{Performance of CTL for Sequential Labeling}
\label{sec:ctl-result}

\begin{table}
\centering
\begin{tabular}{r|r|c}
\hline
\multicolumn{2}{r|}{\bf{System}} & \bf{Loc. $F_1$ (\%)} \\
\hline
\multicolumn{2}{r|}{\bf{Strong labeling (topline)}} & 67.38 \\
\hline
\multicolumn{2}{r|}{\bf{MIL (baseline)}} & 55.83 \\
\hline
\multicolumn{2}{r|}{\bf{CTC}} & 31.91 \\
\hline
\multirowcell{3}{\bf{CTL}} & Max concurrence = 1 & 59.92 \\
& Max concurrence = 2 & 57.49 \\
& Max concurrence = 3 & 53.63 \\
\hline
\end{tabular}
\caption{Localization performance of the four systems.}
\label{table:performance}
\end{table}

Table~\ref{table:performance} lists the highest evaluation $F_1$
obtained by the various sys-tems within 100~checkpoints.
The CTC system falls long behind the baseline;
as we shall see, this is due to the ``peak clustering'' problem.
The CTL system (with a max concurrence of~1) successfully
outperforms the baseline, and closes a third of the gap
between the baseline of MIL with \pa labeling and
the topline of strong labeling.
A class-wise error analysis shows that the CTL system
exhibits a uniform improvement across classes,
outperforming the MIL baseline for 28 of the 35 event types.
In addition, it appears unnecessary to allow multiple
labels to occur at the same frame.

\begin{figure}[t]
\centering
\includegraphics[width=\columnwidth]{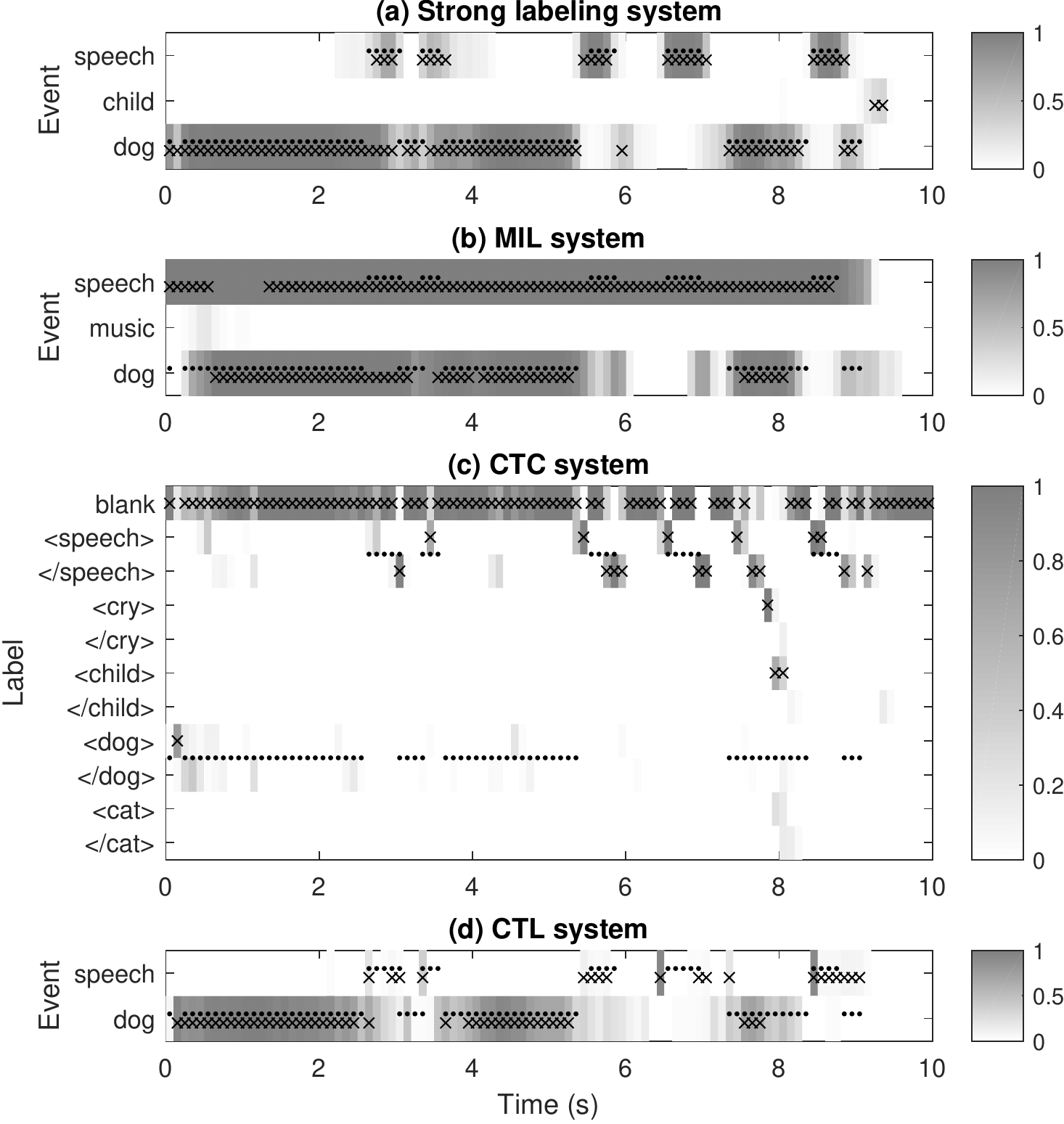}
\vspace{-2em}
\caption{The frame-level predictions of the four systems
on the evaluation recording \texttt{0F04c\_rY4aw}.
Dots stand for the ground truth;
shades of gray indicate the frame-level probabilities of
events, event boundaries or the blank label.
Crosses indicate the most probable label at each frame
(for the CTC system), or events with probabilities higher than
the class-specific thresholds (for the other systems).
\texttt{<E>} and \texttt{</E>} stand for the onset and offset labels
of the event \texttt{E}.
Unimportant events are omitted.}
\label{fig:examples}
\end{figure}

Fig.~\ref{fig:examples} presents the output of the four systems
on an evaluation recording,
which contains the whining of a dog intermingled with speech.
The topline strong labeling system localizes both events well;
the baseline MIL system fails to localize the \texttt{speech} event.
The CTC system can localize the occurrences of \texttt{speech}
(although with a few spurious detections);
for the \texttt{dog} event, however, it exhibits the
``peak clustering'' problem:
it predicts (with low confidence) many pairs of
onset and offset labels of \texttt{dog} next to each other.
The CTL system avoids the ``peak clustering'' problem,
and also localizes the \texttt{speech} occurrences better
than the MIL system.

\subsection{Combining Sequential Labeling with \PA Labeling}
\label{sec:combine}

\begin{figure}[t]
\centering
\includegraphics[width=0.96\columnwidth]{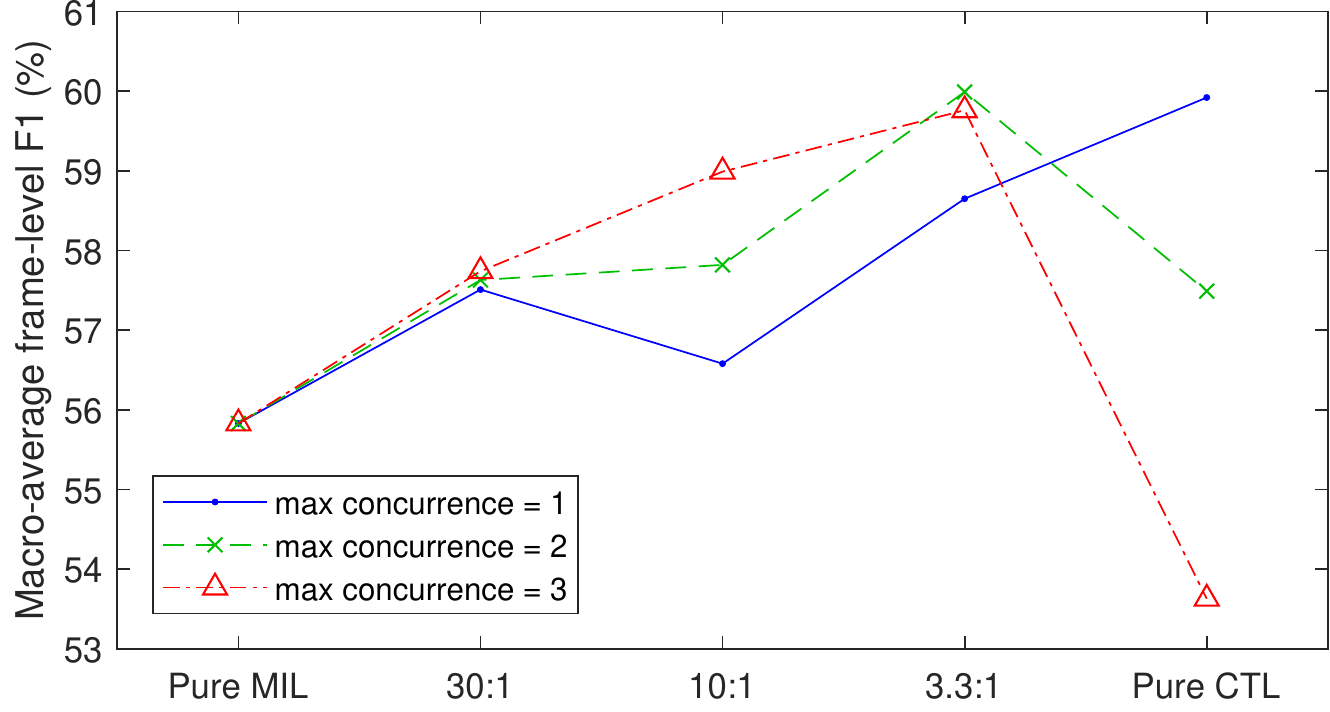}
\caption{The localization performance obtained by
combining CTL and MIL with different weights.}
\label{fig:combine}
\end{figure}

When sequential labeling is available for training a SED system,
\pa labeling is automatically also available.
This prompts us to think about combining a CTL system trained with
sequential labeling and an MIL system trained with \pa labeling.
Because the two systems share all layers up to the
frame-wise probabilities of events,
this combination turns out to be surprisingly easy:
it suffices to combine the loss functions of the two systems
using a weighted average.
At test time, the localization output can be directly taken
from the shared layer of frame-wise event probabilities.
In contrast, it is more difficult to combine a CTC system
with an MIL system because they have different output ends.

We combined an MIL system with CTL systems trained with
different values of max concurrence: 1, 2 and 3.
When we trained the systems alone,
we found that the loss of the CTL systems usually stabilized around 0.2,
while the loss of the MIL system usually stabilized around 0.02.
For the combination experiments, we fixed the weight of the CTL loss to 1,
and tried out the following weights for the MIL loss:
30 (emphasizing the MIL loss more),
10 (weighting both losses equally), and
3.3 (emphasizing the CTL loss more).
The resulting localization performances are plotted
in Fig.~\ref{fig:combine}.
A mixing weight of 3.3:1 appears to be generally a good choice,
and gives a marginal improvement on top of pure CTL.

The potential use of combining a CTL system with other systems
is not limited to the experiments above.
Because sequential labeling takes more effort to produce
than \pa labeling after all, it can be well imagined that
there will be less data with sequential labeling available
than data with \pa labeling.
System combination allows us to exploit the information
in both types of labeling:
we can compute the MIL loss on all the data
and the CTL loss on the part of the data with sequential labeling,
and train a system to minimize an appropriate weighted average
of the two loss functions.
If we also have data with strong labeling,
then the frame-wise cross-entropy loss of a strong labeling system
can be added to the weighted average, too.
A CTL system can be combined with an MIL system and
a strong labeling system with no effort,
thanks to the fact that it computes frame-wise probabilities of events
in the same way as the other two systems.

\section{Conclusion and Discussion}
\label{sec:conclusion}

We made three modifications to the connectionist temporal classification (CTC)
framework: (1) instead of predicting frame-wise boundary probabilities directly,
the network predicts event probabilities and then derives boundary probabilities
using a ``rectified delta'' operator;
(2) the boundary probabilities at the same frame are regarded as
mutually independent instead of mutually exclusive;
(3) the mapping $\mathcal{B}$ from alignments to unaligned label sequences
no longer collapses consecutive repeating labels.
The resulting framework, which we name ``connectionist temporal localization'' (CTL),
successfully solves the ``peak clustering'' problem of CTC,
and closes a third of the gap between the baseline of \pa labeling
and the topline of strong labeling.

Because a CTL system predicts frame-wise event probabilities
in the same way as an MIL system for \pa labeling and
a strong labeling system,
the combination of the three systems is as easy as a weighted average
of the loss functions.
This makes it possible to exploit the information
in all three types of labeling when we have different data
labeled at different granularities.

For more details about the CTL algorithm and the experiments,
please refer to Chapter~4 of the first author's PhD thesis~\cite{Yun-PhD-Thesis}.
The code and acoustic features for the experiments are available at
{\footnotesize \url{https://github.com/MaigoAkisame/cmu-thesis}}.

\vfill\pagebreak

\section{REFERENCES}
\label{sec:refs}

\printbibliography[heading=none]

\end{document}